# A Product Line Systems Engineering Process for Variability Identification and Reduction

Mole Li, Alan Grigg, Charles Dickerson, *Senior Member, IEEE* , Lin Guan and Siyuan Ji, *Member, IEEE*

*Abstract*— **Software Product Line Engineering has attracted attention in the last two decades due to its promising capabilities to reduce costs and time to market through reuse of requirements and components. In practice, developing system level product lines in a large-scale company is not an easy task as there may be thousands of variants and multiple disciplines involved. The manual reuse of legacy system models at domain engineering to build reusable system libraries and configurations of variants to derive target products can be infeasible. To tackle this challenge, a Product Line Systems Engineering process is proposed. Specifically, the process extends research in the System Orthogonal Variability Model to support hierarchical variability modeling with formal definitions; utilizes Systems Engineering concepts and legacy system models to build the hierarchy for the variability model and to identify essential relations between variants; and finally, analyzes the identified relations to reduce the number of variation points. The process, which is automated by computational algorithms, is demonstrated through an illustrative example on generalized Rolls-Royce aircraft engine control systems. To evaluate the effectiveness of the process in the reduction of variation points, it is further applied to case studies in different engineering domains at different levels of complexity. Subject to system model availability, reduction of 14% to 40% in the number of variation points are demonstrated in the case studies.**

*Index Terms*— **Product Line, Relational Orientation, Variability Modeling, Systems Engineering, Systems Modeling Language.**

## I. Introduction

THE advances of Software Product Line Engineering (SPLE) during the last two decades to improve time to market, cost, quality and productivity are primarily attributed to identifying commonalities and variabilities between existing systems that provide greater reusability [1]. Systems Engineering has also recognized the need for reusability [2]. However, applying state-of-the-art SPLE methods [3]-[6] to system domains in addition to the software domain is not straightforward. Thus, extending models and methods for SPLE to the broader domain of Systems Engineering offers new challenges. Our previous work [7][8] addressed these challenges by providing a practical variability modeling method including variability modeling at the requirements engineering and hardware design stages.

Apart from focusing on variability modeling and presentation to include aspects of hardware, current research on extending SPLE to Systems Engineering has also considered automated generation of variability models [9][10]. These approaches have reduced upfront investment and intensive domain analysis required in the development of product line models [9]. However, most of the research in this direction has concentrated on automation of variability identification without considering whether variability should exist in the first instance. As such, the rationale for defining variability in these approaches has not been explicitly and effectively captured. Therefore, understanding the implications of variability configurations and dependency has remained implicit [11]. Furthermore, when combining with hardware variability, overall variability can be ubiquitous in a product line. In a large-scale system, variability configuration can then become complex and even unfeasible [12]. As reported in [13], a modern premium-class car could possibly have up to $10^{20}$ different configurations. Therefore, to keep the configuration process manageable, variability should be kept at a reasonable level.

Motivated by the lack of methods in reducing variability in a variability model, this paper proposes a Product Line Systems Engineering (PLSE) process that offers a novel approach to reduce the number of variation points presented in a given variability model; thereby, reducing the configuration complexity in Product Line Engineering (PLE). The process provides two capabilities: (i) It consumes functional artifacts represented in models developed in the Systems Modeling Language (SysML) to form a unified variability model with hierarchical representation. This is achieved through formalizing dependencies existed in the variability model, the functional artifacts and between the two. (ii) It uses relational transformation as defined in [14] to combine variation points that satisfy completeness and uniqueness criterions. Both capabilities are automated through provision of high-level algorithms. The advantages and limitations of the approach presented in the PLSE process are also assessed by using evaluation cases from different engineering domains.

This work was supported in part by the Rolls-Royce Controls and Data Service team.

Mole Li and Lin Guan are with the Department of Computer Science, Loughborough University, UK(e-mail: m.li@ lboro.ac.uk; l.guan@ lboro.ac.uk).

Charles Dickerson and Siyuan Ji are with the Wolfson School of Mechanical, Electrical and Manufacturing Engineering, Loughborough University, UK(e-mail: c.dickerson@ lboro.ac.uk; s.ji@ lboro.ac.uk).

Alan Grigg is with the Department of Control and Data Services, Rolls-Royce, UK(e-mail: alan.grigg@controlsdata.com).

This paper is structured as follows: Section II provides brief background knowledge for the current advances in SPLE and illustrates key challenges in SPLE with an example of aircraft engine control system. Section III addresses the discussed challenges by proposing the PLSE process. In particular, Section III-(a) introduces the proposed overarching process and is followed by two subsections that elaborate the process in detail. Then, section III-(d) automates the main parts of the overarching process by providing computational algorithms. Section IV evaluates the proposed process through applications to several case studies across various domains and at different levels of complexity. Performance of the PLSE process is discussed in this section as well. Finally, the conclusion and future work are presented in Section V.

## II. BACKGROUND AND MOTIVATION

### A. Current Advances in SPLE

Variability representation techniques enable the modeling of system variability. However, these techniques often rely on experts' experience and require huge upfront human efforts to develop product line models through manually identifying variability. There are several research efforts on automating the identification of system commonality and variability. Niu et al. [9] proposed a method to automatically extract functional requirements from existing textual documents. They adopted Orthogonal Variability Modeling (OVM) approach to represent the identified variability. In [10], the authors presented a method based on Common Variability Language to automatically analyze commonality of system models. Similarly, paper [15] proposed a feature-oriented approach to automatically mine legacy system models via comparing, matching and merging artifacts. Recently, [16] proposed an approach that reuses Domain Specific Languages models and synthesizes with the corresponding OVM variability model. K. Czarnecki et al. [17] proposed a feature models mining method based on conjunctive and disjunctive association rule mining. They proposed the probabilistic feature models (PFM) to represent the mining results. PFM facilitates variability configuration via adding probabilities to certain selections. However, the accuracy of probabilistic calculation is highly dependent on the number of existing data. These related works discussed above, attempted in tackling the challenge of the expensive upfront cost in SPLE via mining variability elements. However, the presented approaches have the following limitations: insufficient consideration of domain knowledge, lack of automated identification of dependencies between variability, and limited scope to only consider software domain.

Modeling dependencies between variabilities can provide promising value to PLE. According to [18][19], they highlighted several critical requirements for modeling in industries. These include modeling variability as first class elements; enabling hierarchical structure representation in variability model; and especially, modeling dependencies between variabilities. M. Sinnema et al. [19] and S.Bühne et al. [20] both provided detailed classification of variability dependencies. However, these works mainly considered software domain and do not offer an automated method to identify variability dependencies.

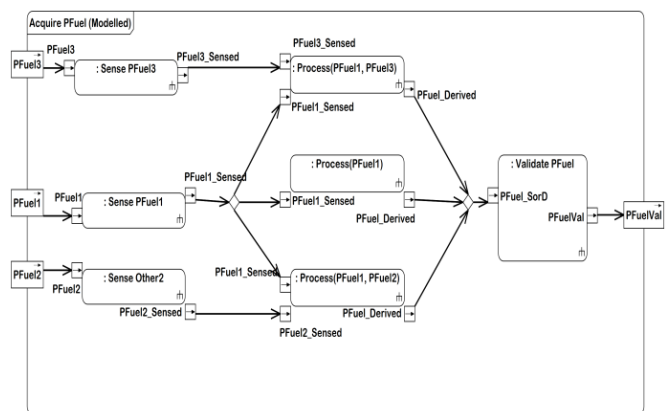

Fig. 1(a) System variability modeled in SysML Activity diagram.

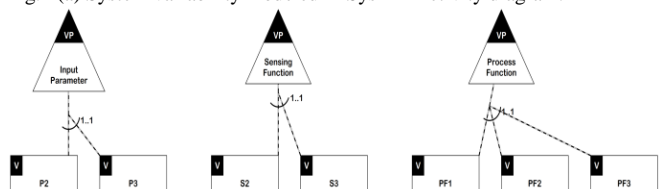

Fig. 1(b) Variability model derived based on Fig. 1a).

### B. Challenges in SPLE

To illustrate the key challenges, including variability dependency analysis and variability configuration in SPLE that will be addressed by the proposed PLSE process, a simplified model of a Rolls-Royce aircraft engine control system, as depicted in Figure 1 will be used in this section.

Rolls-Royce is an international corporation that manufactures jet engines for commercial airlines and military aircrafts. In Figure 1(a), a SysML Activity diagram is depicted to demonstrate an example of combining variable designs of a system into one model. By analyzing commonality and variability of this system, domain engineers can define input parameters, sensing functions and processing functions as variation points. These variation points are depicted in the variability model in Figure 1(b). As shown, the following variants are defined: Input Parameter P2 (for "PFuel2") and P3 (for "PFuel2"); Sensing Function S2 (for "Sense PFuel2") and S3 (for "Sense PFuel3"); and Processing Function PF1 (for "Process(PFuel1)"), PF2 (for "Process(PFuel1, PFuel2)"), and PF3 (for "Process(PFuel1, PFuel3)"). The input parameter PFuel1 and the sensing function "Sense PFuel1" are not considered as variants in the variability model. This is because that they are considered as mandatory system elements in this example. Based on the variability model constructed, without introducing any variability dependency, one can generate $2 \times 2 \times 3 = 12$ different products according based on the given cardinality, i.e. choosing one and only variant for each variation point. Examples of generated products via configuration include {P2, S2, PF2}, {P2, S3, PF1}, and {P2, S3, PF1}. Most of these results (without introducing variability dependency) are invalid designs according to knowledge that are derived from Systems Engineering models. For example, the product {P2, S3, PF1} is invalid because there is no data flow between P2 and S3. From this simple example, it is evident that introducing variability dependency can provide significant value in reducing the configuration space of a product line model. However, manually identifying these dependencies can be difficult and likely requires high upfront investment in large scale models where variability can exist in a multi-level hierarchy. A means for automatic dependency identification and analysis hence becomes demanding.

Although introducing variability dependencies can reduce the configuration space, they bring an additional challenge in reducing configuration complexity. To see where this challenge comes from, the following formal definitions for variability configuration are proposed using the set builder notation. Also note that indices, e.g. $i, j$, do not possess global meanings; they are used to either demonstrate local consistency or distinguish elements under the same set. As such, different index systems can be used in practice as long as Definitions are satisfied.

**Definition** 1 A Product Line Model is defined as a 5-tuple $(VP, V, A, \Delta, \Theta)$ [21] where:

- Let $VP$ be the set of *variation points*, $vp_i : vp_i \in VP$. Each variation point, $vp_i$ defines where individual variability occurs.

- Let $V_i : V_i \in V$ be a set of *variants*, $v_i : v_i \in V_i$, that contains all the possible options of the variation point $vp_i$.

- Let $A$ be a set of *functional artifacts*, $a_i : a_i \in A$, in which $a_i$ is a set of *activities, $act_i : act_i \in a_i$.

- $\Delta$ is a *realization dependency* such that the set of variants, $V_i$ realizes the corresponding variation point denoted by $vp_i$; hence, $(V_i, vp_i) \in \Delta$. Each individual variant, $v_i$ also inherits this dependency, i.e. $\forall v \in V_i : ((V_i, vp_i) \in \Delta) \leftrightarrow ((v, vp_i) \in \Delta)$. Note that in the rest of this paper, defined formal relations are always denoted using non-italic capital letters.

- $\Theta$ is an *artifact dependency* that binds directly a functional artifact to a variation point, i.e. $(a, vp) \in \Theta$, and the activities contained in the functional artifact to the corresponding variant, i.e. $(act, v) \in \Theta$ with $act \in a$ and $(v, vp) \in \Delta$.

**Definition** 2 The configuration of Product Line Model can be described as a process [21], $P$, where:

$$P = \forall vp \in VP:$$
$$\left( \exists v \in V, a \in A, act \in a : (\Delta(v, vp) \wedge \Theta(a, vp) \wedge \Theta(act, v)) \right).$$
(1)

This process is explained as follows: for each variation point in $VP$, there is at least one variant, $v$, that binds an activity, $act$, and realizes the $vp$ in which is context is defined by the functional artifact, $a$ that contains the activity.

From Definition 1, it is observed that as long as variability exists, a corresponding variation point can be defined. Then, for every variation point defined, dependency analysis will include the identification of realization dependency, $\Delta$ and artifact dependency, $\Theta$. As such, with increasing number of variability points, as observed in Definition 2, the configuration space grows quickly and eventually becomes unmanageable. Therefore, in order to maintain a manageable complexity of the configuration process, methods that reduce variation points can bring promising value.

Motivated by the potential benefits in tackling these two challenges described above, this paper aims at proposing a process to address these challenges.

## III. THE PLSE PROCESS

To tackle the challenges stated in the previous section, this paper proposes a Product Line Systems Engineering (PLSE) process that utilizes a Relational-Oriented Systems Engineering (ROSE) concept previously developed in [14][22] to: (i) formalize existing dependencies in SysML system models, (ii) to define dependencies between product line models and their related SysML models, and (iii) to reduce variability complexity by utilizing relational transformation.

### A. An Overarching Process for Variability Identification and Optimization

This subsection proposes key definitions and the underlying methods for the PLSE process. The overarching process is depicted in Figure 2. There are four Stages in the process.

Stage 1 is concerned with the reuse of legacy system models to generate a three-layer hierarchical model structure. This stage begins with importing *functional artifacts,* which are defined as system functional models that are represented in modeling languages such as SysML. In particular, the functional artifacts considered in this paper are referred to system functional models presented in SysML Activity Diagrams. We note here that the scope of this paper is restricted to functional artifacts only. Other system artifacts such as physical artifacts will be a topic of future research. Based on the imported functional artifacts, a new functional artifact with a three-layer structure is generated at the end of Stage 1. The three layers are: (i) the *Feature Layer*. Different from the concept of "Feature" in Feature Modeling method [23], the Feature Layer is the highest-level abstraction used to represent the abstraction of groups of system functional requirements. This is to provide a high-level reusability. (ii) The *Functional Layer*. This layer presents the details of the system functional requirements that are often modeled in SysML Requirement diagram. However, within the scope of this work, the Functional Layer is derived based on SysML Activity models in which each system functional requirement is modeled as an Activity (denoted as *fc*). The reason for using Activity model is that it does not only provide requirements information on system functional requirements, but also relationships among them. (iii) The *Component Layer*. This layer consists of design models that represent system functional designs. These models can be transformed into low-level software requirements for the software design team. The above hierarchical approach has been developed and adopted by the Rolls-Royce Control and Data Services team who are supporting the current research. The details of this sowfware level approach are beyond the scope of this paper. Based on the three-layer definition, if legacy system functional requirements are available in the form of SysML Activity models, the Functional layer and the Component Layer are considered to be pre-defined. Hence, only the abstraction of these SysML Acitivity models into Feature Layer (Step 1.2) is required to complete Stage 1. However, if available system functional requirements in terms of Activity diagram are in alternative forms such as the covential SysML Requirement models, Step 1.3 is required to first abstract the Component Layer information into the Functional Layer in accordance with the functional requirements defined in the available functional artifacts.

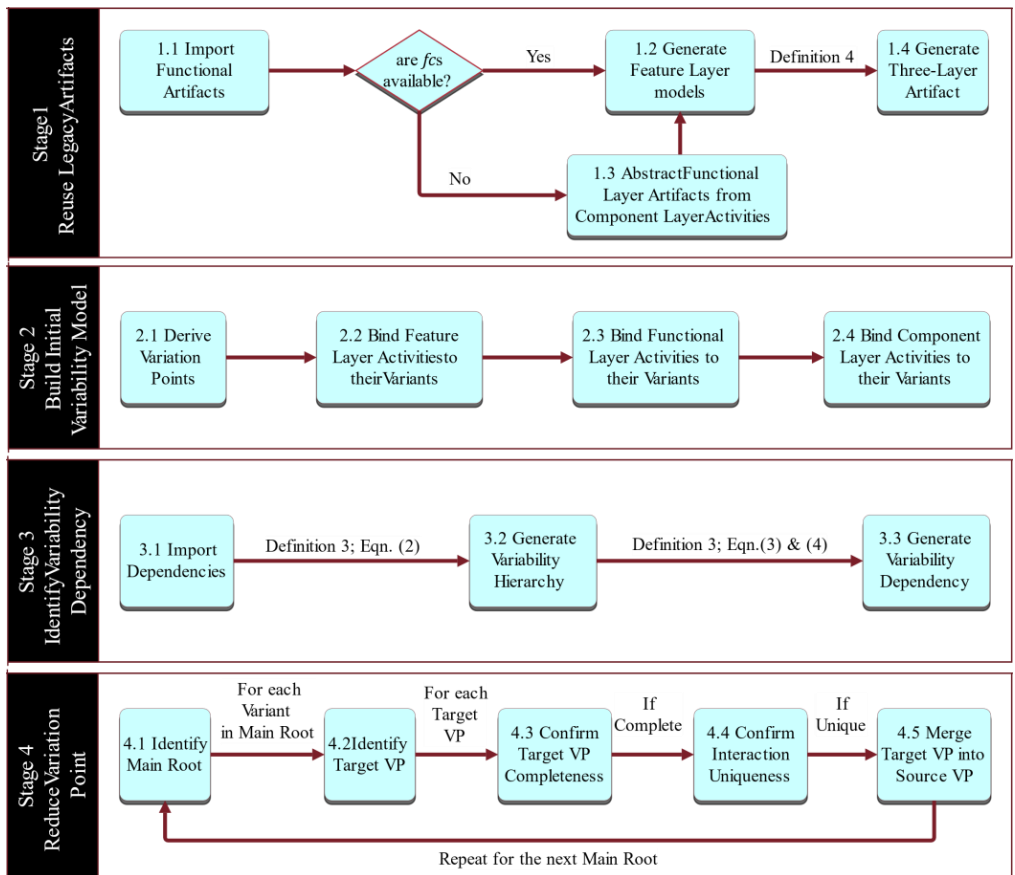

Fig. 2.  The proposed ROPLSE overarching process diagram. (VP is short for Variation Point)

Stage 2 is concerned with the derivation of an initial variability model based on the generated three-layer functional artifact. variation points are first created in Step 2.1 based on variability identified in the three-layer functional artifact. Step 2.2 through Step 2.4 then defines corresponding variants and variation points for each of the three-layers and their artifact dependencies (c.f. Definition 1).

Stage 3 is concerned with the preservation of the relational structure of the three-layer functional artifact to the initial variability model. It refers to identifying variaility dependencies between variants based on relational structure of the functional artifacts. The relational preservation is fulfiled by using iterations between activities in the functional artifacts within the same layer. For example, a dependency between two functional requiremsnts together with the bindings from the functional requirements to the corresponding variants are used to identify dependencies between variants through relational transformation in Step 3.2. Details of this transformation is explained in the next subsection. Similarly, dependencies identification is also applied to the other two layers to transform relations into the initial variability model.

Finally, Stage 4 is concerned with optimizing the initial variability model through a reduction in the number of variation points. The stage starts with identifying the *Main Root*. Then, going through a sequence of steps (Step 4.2 to Step 4.5), the reduction is achieved by *merging* the *target variation point* into the *source vairation point* if the target variation point is confirmed *complete* and that the interactions between the source and the target variation point are *unique*. Detail of this reduction method and definitions of Main Root, source and target variation points, completeness, uniqueness and merging will be explained in Section III-C. The algorithimitc automation of the reduction method will be developed in Section III-D. These steps are repeated for the next Main Root until no further merging can be achieved.

To summarize, the first three stages presented in the PLSE process diagram aim to reuse functional artifacts to create an initial variability model in which the structural information is persevered. The last stage aims to optimize the generated initial variability model by reducing the number of variation points presented in the model.

### B. Variability and Variability Dependency Identification

This section discusses in depth the first three stages depicted in the PLSE process diagram and proposes relevant definitions and methods to enable the elementary processes.

As stated in [24], the formalization of variability modeling language has benefits in avoiding unnecessary misinterpretations. Definition 1 generally represents the core idea of SPLE. However, this definition does not support a hierarchical representation of a variability model. Given that a variability model will be derived from a three-layer functional artifact, it is therefore important for the variability model to capture the following important relationships to capture the relational structure of the three-layer functional artifact. These relationships are: refinement relation in which a variation point can refine a higher-level variant, and dependencies between variants. Definition 3 and Definition 4 below aims at providing a means to capture the above relationships. This is achieved by extending Definition 1 with introducing an elaborated definition for a variability model with hierarchy and an elaborated definition for a three-layer functional artifact.

**Definition** 3 A variability model with hierarchy is a tuple $VM = (VP, V, \Delta, T)$ where the definitions for $VP$, $V$ and $\Delta$ remains unchanged, and $T$ is a set of *variability trees* with $T = (vp^r, \Psi, \Phi)$.

- Let $vp^r: vp^r \subset VP$, be a set of *root variation points,* with $vp^r_i$ being the root variation point of the $i^{\text{th}}$ tree, i.e. the $i^{\text{th}}$ root.

- Let $\Psi$ be the *variability refinement relation* that refines a higher-level variant $v^\uparrow \in V$ as a *parent* by a lower-level variation point $vp^\downarrow \in VP$ and its variants $v^\downarrow \in V$ as *children*, such that $(vp^\downarrow, v^\uparrow) \in \Psi$ and $(v^\downarrow, v^\uparrow) \in \Psi$ with $(v^\downarrow, vp^\downarrow) \in \Delta$.

- Let $\Phi$ be the *variability dependency* that relates two variants. Variability dependency can be further specialized into various types depending on its semantic meaning. For instance, «requires» relation [25] and «exclude» relation [25] are two well established types of variability dependencies.

**Definition** 4 A three-layer functional artifact is a tuple $A = (a, act, \Gamma, I)$ where:

- Let $a$ be a set of *activities, $act_{ij}: act_{ij} \in a_i$. Based on the three-layer hierarchical structure of the functional artifacts, three types of specialized artifacts and activities are further specified. Firstly, for the Feature Layer, let $F: F \subset a$ be the set of *Feature Layer artifacts* in which each *feature*, as represented by an activity, is denoted by $f_i: f_i \in F$. Secondly, a set of *Functional Layer artifacts* is denoted by $FC: FC \subset a$, with each set, $FC_i$, contains a set of *functional requirements* being denoted as $fc_{ij}: fc_{ij} \in FC_i$. Lastly, for the Component Layer, let $C: C \subset a$ be the set of *Component Layer artifacts* that each contains activities modeling the corresponding *functional designs, $c_{ij}: c_i \in C_i$.

- Let $\Gamma$ be the *activity refinement* presented in the functional artifact $A$. It consists of the following relations. The *feature refinement relation* $\Xi_f: (FC_i, f_k) \in \Xi_f, \Xi_f \subset \Gamma$ is defined as a Functional Layer artifact, $FC_i$ refines a feature, $f_k$. Functional requirements, $fc_{ij} \in FC_i$ inherit this feature refinement relation. Similarly, the *functional refinement relation*, $\Xi_{fc}: (C_i, fc_k) \in \Xi_{fc}, \Xi_{fc} \subset \Gamma$ is defined as a Component Layer artifact, $C_i$ refines a functional requirement, $fc_k$. Functional designs, $c_{ij} \in C_i$ inherit this functional refinement relation. The mapping, $\Gamma \rightarrow \Psi$, through relational transformation provides the Systems Engineering knowledge for where necessary refinement, $\Psi$, should take place in the variability model, $VM$

- Let I be a set of *interactions* between activities, i.e. $(act_i, act_j) \in$ I. Interaction, in software engineering, is generally limited to information exchanges. In Systems Engineering, it involves the transport of energy, material and information between systems [26]. Therefore, interactions between system functions can be specialized into material interaction, MI and information interactions, II . One can then establish the mathematically relation, I = MI ∧ II.

Based on Definition 3 and 4, a Product Line Model can be redefined as a tuple $(VM, A, \Theta)$, where the definition for the artifact dependency, $\Theta$, remains the same. We also emphasize that although the definition for functional artifact is specialized to the three-layer model structure, it can be easily tailored to a system structure that has an arbitrary number of layers that is greater than one.

In the rest of this subsection, along with Figure 3, we demonstrate how the ROSE concept is used in these definitions to generate a variability model with hierarchy. ROSE was first introduced in [22]. It is a general system methodology that offers a coherent mathematical foundation and facilitates discovering and analyzing relations captured by Systems Engineering.

A relational transformation, as defined in [22], is "an association between the elements or parameters of two models

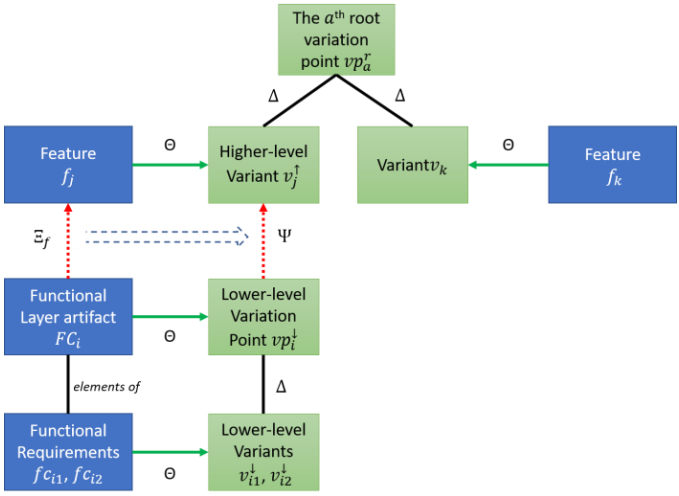

Fig. 3(a). Mapping the hierarchical structure to the variability model.

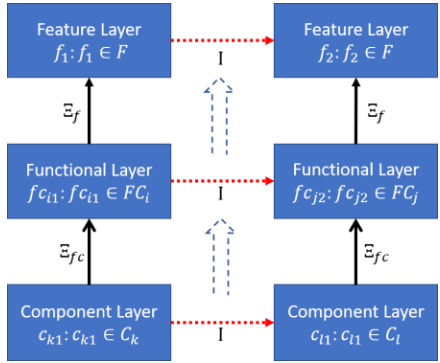

Fig. 3(b). Preserving an interaction through different layers.

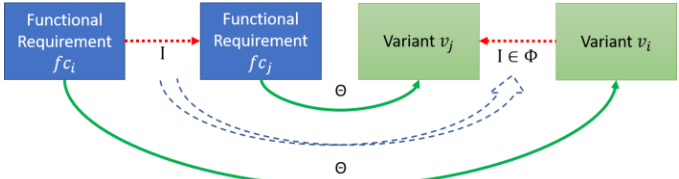

Fig. 3(c). Transforming an interaction to the variability model.

of a system that induces a further mapping between the relationships in the models". To see how a relational transformation is applied to preserve an existing relation, let us consider the following scenario. As depicted in Figure 3(a), a lower-level (e.g. Functional Layer) artifact can refine a higher-level activity (e.g. a Feature in a Feature Layer) through the activity refinement shown in a dotted arrow (e.g. feature refinement relation, $\Xi_f$). If both the higher-level and the lower-level activities are derived into variant and variation points respectively through the artifact dependency, $\Theta$, then, the relational transformation states that the above relations induce (dashed wide arrow) a new relation between the derived variant and the variation point, which is defined as the variability refinement relation, $\Psi$. The above transformation, using the example depicted in Figure 3(a), is captured by the mathematical expression:

$$(FC_i, f_j) \in \Xi_f \text{ with } (FC_i, vp_i^\downarrow), (f_j, v_j^\uparrow) \in \Theta$$

implies $(vp_i^\downarrow, v_j^\uparrow) \in \Psi,$ (2)

with $(v_j^\uparrow, vp_a^r) \in \Delta$, $(v_k, vp_a^r) \in \Delta$, $(v_{i1}^\downarrow, vp_i^\downarrow), (v_{i2}^\downarrow, vp_i^\downarrow) \in \Delta$, $(f_k, v_k) \in \Theta$, $fc_{i1} \in a_i$ and $fc_{i2} \in a_i$ capturing the rest of the relations shown in Figure 3(a).

Using the same ROSE concept, we next demonstrate how it is applied to preserve interactions between activities from a lower-level layer to a higher-level layer in the artifact hierarchy. As depicted in Figure 3(b), the two functional designs $c_{k1}$ and $c_{l1}$ interact via material flow or information flow such that $(c_{k1}, c_{l1}) \in I$. Given that each of the functional designs refines a higher-level functional requirement, $fc_{i1}$ and $fc_{j1}$ respectively via the functional refinement relation, $\Xi_{fc}$, the interaction between the two functional designs induces an interaction between the two functional requirements, such that $(fc_{i1}, fc_{j1}) \in I$. This relational transformation is mathematically expressed as,

$$(c_{k1}, c_{l1}) \in I \text{ with } (c_{k1}, fc_{i1}), (c_{l1}, fc_{j2}) \in \Xi_{fc}$$

implies $(fc_{i1}, fc_{j2}) \in I.$ (3)

Similar to Figure 3(a), the dependency of interest, I, is depicted using dotted arrow and the transformation is illustrated using dashed wide arrow in Figure 3(b). In addition, as shown in Figure 3(b), the relational transformation can be further applied to induce an interaction between two features in the Feature Layer.

Finally, we apply the relational transformation to interactions between activities in the three-layer artifact to induce an interaction, I between two variants in the variability model. Without going through the details, the relational transformation for inducing this interaction is mathematically captured by,

$$(fc_i, fc_j) \in I \text{ with } (fc_i, v_i), (fc_j, v_j) \in \Theta$$

implies $(v_i, v_j) \in I,$ (4)

and graphically depicted in Figure 3(c). In addition, based on Definition 3, the induced interaction between variants is then considered as a specialized type of variability dependency, i.e. I ∈ Φ. It is worth noting that interactions identified between variants may or may not be «requires» relations. To appropriately determine the exact semantics of these interactions, additional knowledge is likely required. However, a «requires» relation can always be regarded as an interaction due to the generality of the definition of an interaction.

To illustrate the use of this transformation to identify interactions between variants in practice, we continue with the example given in Section II (Figure 1). As depicted in Figure 1(a), in the top activity flow, the first interaction observed is a material flow from "PFuel3" to "Sense PFuel3". Hence, we write (PFuel3, Sense PFuel3) ∈ I according to Definition 4. As "PFuel3" binds with the variants, P3, under "Input Parameter" variation point, and "Sense PFuel3" binds with the variants, S3, under "Sensing Function" variation point, we have, $(\text{PFuel3}, \text{P3}) \in \Delta$ and $(\text{Sense PFuel3}, \text{S3}) \in \Delta$ according to Definition 1. Then, applying Equation (4), the relational transformation induces an interaction between P3 and S3, i.e. $(\text{P3}, \text{S3}) \in \text{I}$, which is depicted in Figure 4(a). Using the same concept, the following interactions have been identified and captured in Figure 4(a):

$$\begin{cases} (\text{S3}, \text{PF3}) \in \text{I} \\ (\text{P2}, \text{S2}) \in \text{I} \\ (\text{S2}, \text{PF2}) \in \text{I} \end{cases}. \quad (5)$$

The induced variability dependencies suggest that the selection of the variant, "PF1" does not have impact on any other variabilities. This is explained by the fact that its binding activity, "Process(PFuel1)", is only related to the mandatory system element, "Sense PFuel1".

### C. Variation Points Reduction

Variability reduction stage starts with consuming the variability model generated by the end of Stage 3. As illustrated in Figure 2, the first step (Step 4.1) in this stage is to identify the *Main Root*, which is defined as the variability tree (as denoted by its root variation point, $vp_t^r$) that contains the largest number of variants in the tree. The following example demonstrates how the number of variants in a variability tree is calculated. Consider a $vp_t^r$ that has 2 variants A and B with A refined into 5 lower-level variants (in here we ignored variation points-variant structure for simplicity), B refined into 3 variants. Then, the total number of variants for this tree is calculated as $2 + (5 + 3) = 10$ variants.

After the Main Root is identified, for each variant, $v_i$ in the Main Root, the process traces its interactions to all other variants, $v_j$ that are at the same level but realize other variation points. Mathematically, this is expressed as $(v_i, v_j) \in \text{I}$ or $(v_j, v_i) \in \text{I}$ with $(v_i, vp_s) \in \Delta$ and $(v_i, vp_t) \in \Delta$, where $s \neq t$. Also note that as long as interactions are identified based on Equation (4) in Stage 3, variants that are interacting are guaranteed to be at the same level. Then, through these traces, two variation points are considered as interacting if their realizing variants are interacting. In the pair of interacting variation points, the variation point that has a relatively smaller number of variants (compared in the same level) is defined as a *target variation point* $vp_t$; and the variation point with larger number of variants is defined as the corresponding *source variation point*, $vp_s$. With the above, all interacting variation points are identified in Step 4.2. We also emphasize that the convention of 'source' and 'target' is independent of the direction of the interaction.

Then for each target variation point, the process will confirm its completeness in Step 4.3 and then uniqueness of the relevant interactions in Step 4.4.

*Completeness* is defined as a situation where for each variant that is realizing the target variation point, there is at least one interaction between this variant and a variant that is realizing the corresponding source variation point. Mathematically, this is expressed as,

$$\forall v_j \in V_j:$$
$$\Big( (v_j, vp_t) \in \Delta, (v_i, vp_s) \in \Delta, \exists \big( (v_i, v_j) \lor (v_j, v_i) \big) \in \text{I} \Big). \quad (6)$$

Physically, if completeness is satisfied, it means that the target variation point has a strong dependency with the source variation point in which the option of any variant for the target variation point is related to the configuration of the source variation point.

Once completeness is satisfied, the process continues to confirm *uniqueness*, which is defined as follows: An interaction is *unique* if and only if it is the only interaction path between the two variants; if all interactions involved in the completeness checking are unique, then it is said that the target variation point satisfies uniqueness. Similar to how paths are defined in graph theory, an interaction path between two variants is defined as a sequence of interaction(s) connecting the two variants via a sequence of other variant(s). Note a direct interaction path involves no other variants. An example of a non-unique interaction is mathematically expressed as:

$$\forall (v_i, v_j) \in \text{I}: \Big( \exists \big( (v_i, v_k) \in \text{I} \land (v_k, v_j) \in \text{I} \big) \Big). \quad (7)$$

Once uniqueness is confirmed between a pair of source and target variation points, the target variation point is merged into the source variation point in Step 4.5. In this step, *merging* has three aspects: (i) the target variation point at the current level will be removed; (ii) to keep the original information on artifact dependencies, the activities that are originally bound to the removed variants are now bound to the variants under the source variation point. (iii) The lower level variation points (and the rest of the tree) that originally refine the target (removed) variant are now absorbed into the corresponding variants under the source variation point. Mathematically, merging $vp_t$ with $(v_j, vp_t) \in \Delta$ into $vp_s$ with $(v_i, vp_s) \in \Delta$ means removing $v_j$, $\big( v_j^\dagger, vp_j^\dagger \big) \in \Psi$, $vp_t$ and $(act_j, v_j) \in \Theta$ for all $j$; and add $(act_j, v_i) \in \Theta$ and $\big( v_i^\dagger, vp_j^\dagger \big) \in \Psi$ for every $(v_j, v_i) \in \text{I}$ or $(v_j, v_i) \in \text{I}$ to the Product Line Model.

The reduction stage continues iteratively by repeating from Step 4.1 to Step 4.5 with the identification of the next Main Root.

In the following, the introduced method for the reduction of variation points is illustrated through continuing the example in Section III(b). After the variability model (output of Stage 3) is generated and as depicted in Figure 4(a), the Main Root, "Process Function" has been identified. It has 3 variants that each binds with a corresponding activity in Figure 1(a). Based on Step 4.2, the "Sensing Function" is first identified as a target variation point due to the interactions S2 to PF2 and S3 to PF3. Since S2 and S3 are the only two variants in the "Sensing Function", and both of them are interacting with a variant in the "Process Function", this target variation point is considered complete (Step 4.3). Then, as both S2 to PF2 and S3 to PF3 are unique, the "Sensing Function" target variation point is also

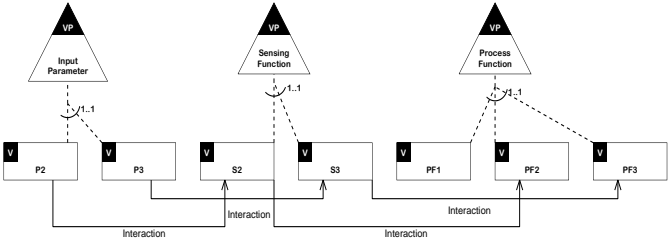

Fig. 4 (a). Dependency identification and annotation.

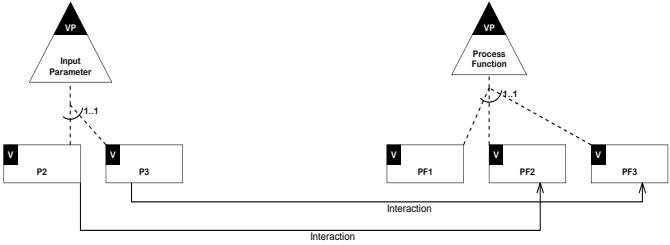

Fig. 4 (b). Merging a variation point.

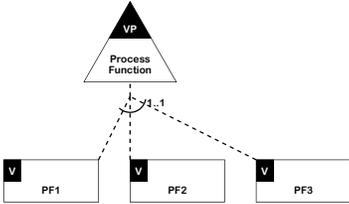

Fig. 4 (c). The optimized variability model.

confirmed unique (Step 4.4). Therefore, this variation point can be merged into the "Process Function" source variation point by removing the "Sensing Function" variation point and re-binding the relevant activities (Step 4.5). The resultant variability model at this point is depicted in Figure 4(b). Note that the interactions, P2 to S2 and P3 to S3 are transferred in the merging process and now read as P2 to PF2 and S3 to PF3. The reduction continues by repeating from Step 4.1 to Step 4.5, in which the next Main Root is identified as "Input Parameter". Following the same arguments, "Input Parameter" is merged into the "Process Function" since both completeness and uniqueness are satisfied. The final, optimized variability model is depicted in Figure 4(c). Stage 4 ends here as no new Main Root is identified.

Finally, we illustrate a non-unique interaction path using the above example. Let us imagine a direct interaction path from P3 to PF3 (not depicted in Figure 4(a)). This imagined interaction path would not be a unique one based on Equation (7), because it is not the only interaction path from P3 to PF3. The other interaction path is P3 to S3 to PF3 as depicted in Figure 4(a).

### D. Algorithms

This section develops algorithmic approaches to automate the functionalities proposed in the PLSE process.

Algorithm 1 is developed to computationally semi-automate the process of identifying variation points and creating variability model (Stage 2 and Stage 3 of the PLSE process). It imports a set of three-layer functional artifacts $A$. The output of this algorithm is an initial variability model, $H$ (c.f. Definition 3). Line 3 introduces a comparison function that compares every element of the functional artifacts in each abstraction layer. Where a different model element is identified (Line 4), a $vp$ is created (Line 5). In addition, Line 6 creates a variant, $v$, according to the model difference and binds the variant with the corresponding activity, $dif$ via the artifact dependency, $\Theta$. Note that in the rest of this section for algorithms development, a relation set, e.g. $(dif, v) \in \Theta$ will be denoted in the form of element $x$ relates to element $y$, e.g. $dif\Theta v$.

From Line 8 to Line 19, a function with two input arguments, called 'Mapping', is defined. The two inputs define the two layers (scope) for which the functions will be applied to. In this function, the function considers every $dif$ identified in the layer, $L$ (Line 9). If there is an interaction between a pair of $dif$s (Line 10), then, based on Equation (4), the function creates an interaction between their binding variants, $v_i$ and $v_j$ (Line 11). The function continues to check whether each of the $dif$s refines an activity, $act$ in the input layer, $U$ (Line 12). If the result is positive, then, based on Equation (3), the function creates an interaction between the two activities, $act_k$ and $act_l$ (Line 13). Finally, at Line 14, if the two activities are binding with corresponding variants, i.e. that themselves are also $dif$s, the function utilizes Equation (2) to preserve the hierarchical refinements, $\Gamma$, into the variability model by creating variability dependencies between the relevant variants, $v_i^{\dagger}\Psi v_k^{\dagger}$ and $v_j^{\dagger}\Psi v_l^{\dagger}$ (Line 15 and 16). Based on the formal definition of the three-layer artifact, $A$, the conditional loop from Line 12 to Line 18 will be excited if and only if the second input argument, $U$ is exactly one layer above the first input argument, $L$.

To generate the initial Variability Model, the algorithms call the function three times with different sets of input arguments. In particular, Line 22 applies Mapping from the Component to the Functional Layer; Line 23 applies Mapping from the Functional Layer to the Feature Layer; and Line 24 applies Mapping in the Feature Layer only such that Line 12 to Line 18 are not executed.

Algorithm 2 is developed to computationally automate the variation point reduction mechanism (Stage 4 of the PLSE process). The initial variability model generated from Algorithm 1 becomes the input of Algorithm 2. The output of this algorithm is an optimized variability model. Line 3 calculate the size of every variability tree (denoted by its corresponding root variation point, $vp_n^r$) in the set, $T$, by counting the total number of variants within the variability tree. In a descending order of the size of the threes, as in Line 4, for every variant in the current tree, $v_i : v_i \Delta vp_k$, the algorithm searches for other variants that are realizing a different variation point, i.e. $v_j : v_j \Delta vp_l$. If there is an interaction between $v_i$ and $v_j$, despite the direction of the interaction (Line 6), then, based on Equation (6), the algorithm defines a target and a source variation point, $vp_s$ and $vp_t$, respectively in Line 7 and 8.

---

**Algorithm 1: Initial Variability Model Generation**

1: **Input** $A$
2: **Output** initial Variability Model
3: compare each element of $A$
4: **If** there is difference $dif$ **then**
5:     create $vp$
6:     create $v$ and $dif\Theta v$
7: **End If**
8: **Define Function** Mapping($L, U$)
9:     **For** each $dif \in L$
10:         **If** $dif_i | dif_j$ **then**
11:             $v_i | v_j = dif_i | dif_j$
12:             **If** $dif_i \Gamma act_k, act_k \in U$ **AND**
                    $dif_j \Gamma act_l, act_l \in U$ **then**
13:                 $act_k | act_l = dif_i | dif_j$
14:                 **If** $act_k \Theta v_k$  **AND** $act_l \Theta v_l$ **then**
15:                     $v_i^k \Psi v_i^t = dif_i \Gamma act_k$
16:                     $v_j^k \Psi v_l^t = dif_j \Gamma act_{l,}$
17:                 **End If**
18:             **End If**
19:         **End If**
20:     **End For**
21: **End Function**
22: **Call Function** Mapping($C, FC$)
23: **Call Function** Mapping($FC, F$)
24: **Call Function** Mapping($F, F$)

---

Here, the two functions, Max and Min compare two variations points to determine the variation point with higher number of realizing variants and the one with lower number of realizing variants respectively. In the situation where the number of realizing variants are equal, the first input is assigned as the output of the Max function and the second input is assigned as the output of the Min function.

Then, from Line 9 to 11, the algorithm confirms the following: (i) whether $vp_t$ satisfies completeness (c.f. Step 4.3) and uniqueness (c.f. Step 4.4). These are achieved by checking all variants that realize the target variation point, $v_j: v_j \Delta vp_t$ to confirm whether there is at least one interaction, despite the

---

**Algorithm 2: Variation Points Reduction**

1: **Input** initial Variability Model, $VM$
2: **Output** optimized Variability Model
3:         calculate sizeof($vp_n^r.v$) for all $vp_n^r$ in $T$
4: **For** each $vp_n^r$ **In** $T$ in descending  order of size
5:     **For** each $v_i: v_i \Delta vp_k$ **In** $vp_n^r$
6:         **If** $(v_i | v_j) \vee (v_j | v_i)$ **AND** $v_j \Delta vp_l, k \neq l$ **then**
7:             $vp_s = $ Max($vp_k, vp_l$)
8:             $vp_t = $ Min($vp_k, vp_l$)
9:             **If** all $v_j: v_j \Delta vp_t$  has
                    $(v_s | v_j) \vee (v_j | v_s): (v_s \Delta vp_s)$ **then**
10:                 **If** $v_s | v_j$ is unique for all $v_j$ **then**
11:                     delete $vp_t, v_j \Theta act_j, vp_j^t \Psi v_j, v_j | v_m$
12:                     add $v_s \Theta act_j, vp_j^t \Psi v_s, v_s | v_m$
13:                 **End If**
14:             **End If**
15:         **End If**
16:     **End For**
17: **End For**

---

direction of the interaction, between $v_s$ and $v_j$ with $v_s \Delta vp_s$ (Line 9); and whether each interaction is unique, i.e. it is the only interaction path between the associated two variants (Line 10). Once both statements are satisfied for all $v_j: v_j \Delta vp_t$, the algorithm merges $vp_t$ into $vp_s$ by (i) deleting the original variation point, $vp_t$ (with its realizing variants) in the model, relevant artifact dependencies $v_j \Theta act_j$, relevant variability refinement relations, $v_j \Psi vp_j^t$, and relevant interactions, $v_j | v_m$ (Line 11); and (ii) re-binding the activities to the relevant variants in the source variation point, i.e. $v_s \Theta act_j$, and transferring the variability refinement relations and relevant interactions into the source variation point under the corresponding variant, i.e. $vp_j^t \Psi v_s$ and $v_s | v_m$ respectively (Line 12).

## IV. Case Studies

This section quantitatively evaluates the proposed PLSE process through applications to case studies from different engineering domains and to systems at different levels of complexity.

The first case study is extracted from research [27] which modeled an international logistic company's partial inventory process. The process of the company with its variations was modeled in UML Activity Diagrams. The original OVM will be used for generating the variability model. Feeding the variability model into Algorithm 2, a new variability model is generated in which the "Type of Inventory Records" (TIR) variation point is merged into the "Behaviors of Logistics Execution" (BLE) variation point. The merging of the variation points is explained as follows. As observed in the original variability model (Figure 4 in [27]), each of the two variation points is realized by two variants. The two sets of variants interact with each other through two "requires" relations forming two unique interaction paths, i.e. $(v_{11}, v_{22}) \in R^+, R^+ \in I$ and $(v_{12}, v_{21}) \in R^+, R^+ \in I$ with $(v_{ij}, vp_i) \in \Delta$ and the indices $i = \{1,2\}, j = \{1,2\}$. Then, the two interactions imply that the relation between the two variation points is unique, thereby, allowing them to be merged into one variation point. The merging is also captured in Table I. This shows that the PLSE process can reduce the 5 variation points down to 4 for the inventory process variability model.

TABLE I
EVALUATION CASE STUDY 1 & 2

| Name of Initial VPs | Abbreviation | Merged into |
|---|---|---|
| *Inventory Process of Logistic Company* | | |
| Inventory Accomplishment | IA | |
| Type of Inventory Records | TIR | |
| Base of Inventory | BI | |
| Behavior of Logistics Execution | BLE | TIR |
| Generation of Appointment | GA | |
| *Renault Electric Parking Brake System* | | |
| Parking Brake Service | PBS | |
| Hill Start Assistant | HSA | |
| Gear Box | GB | |
| Regulation Zone | RZ | |
| Vehicle Trailer | VT | |
| Vehicle Properties | VP | |
| Brake Lock | BL | |
| Clutch Pedal | CP | |
| Brake Release | BR | |
| Redundant Functions | RF | |
| HSADisable Function | HF | |
| Architecture Design Alternatives | ADA | |
| Force Distribution Design Alternatives | FDDA | |
| Software Allocation Design Alternatives | SADA | |
| Tilt Angle Function Allocation | TAFA | |
| Force Monitor on Engine Stop | FMES | ADA |
| AUX | AUX | |
| Brake Behavior Thresholds | BBT | BS |
| Braking Strategy | BS | |
| Electrical Action Components | EAC | ADA |
| Input Information | II | |

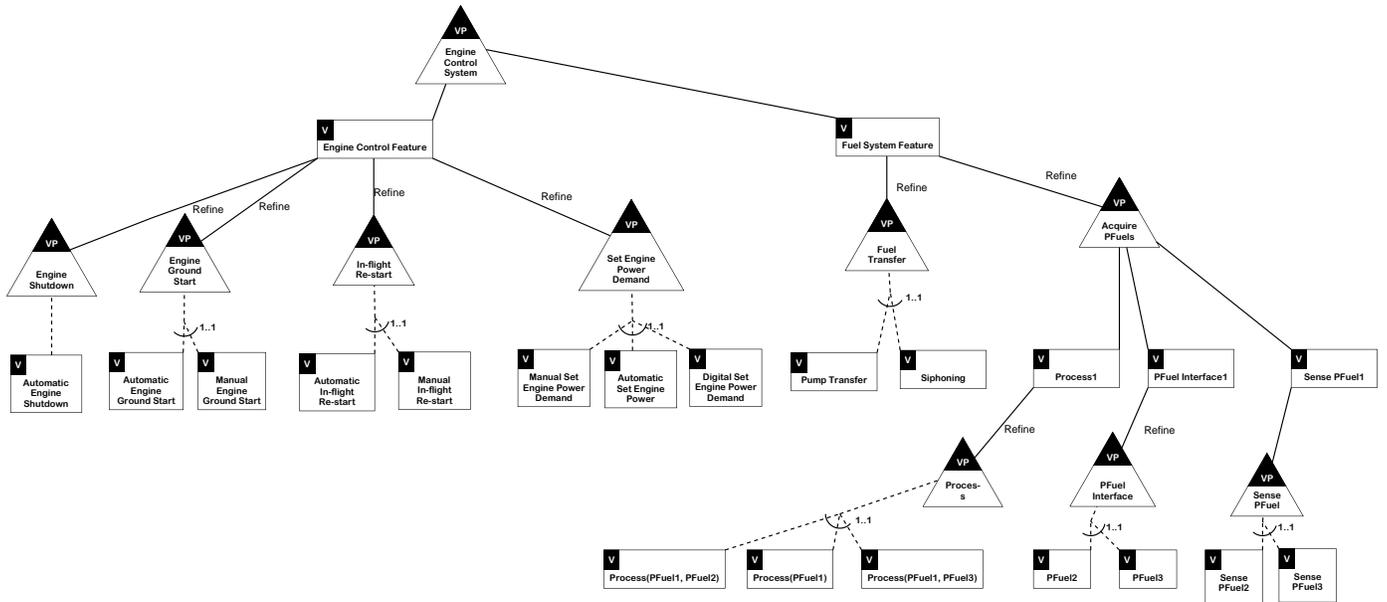

Fig. 5.  Initial variability model for the engine control system in Case Study 3

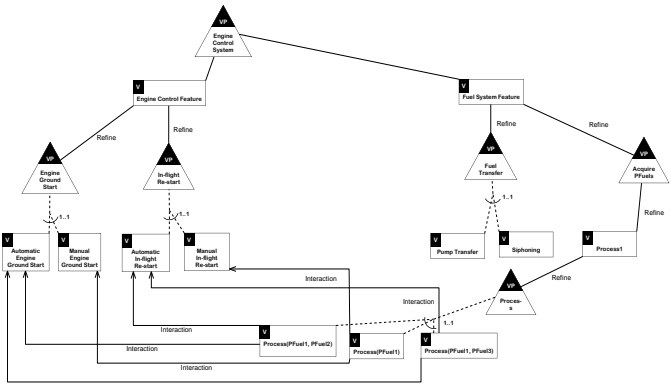

Fig. 6.  Optimized variability model for the engine control system (c.f. Fig. 5).

TABLE II
A SUMMARY OF THE CASE STUDY RESULTS

| Case Study | No. of Initial VPs | No. of VPs after Optimization | Reduction Percentage Domain | Domain |
|---|---|---|---|---|
| 1 | 5 | 4 | 20% | Logistics |
| 2 | 21 | 18 | 14% | Automotive |
| 3 | 10 | 6 | 40% | Aerospace |
| 4 | 37 | 24 | 35% | Aerospace |

The second case study is constructed based in [28], the Renault Electric Parking Brake (EPB) System. The system is commonly used to facilitate drivers' conventional parking brake behavior. The thesis [28] combines the EPB system with "Hill Start Assistant" functionalities for the analysis of system family reuse. To apply the PLSE process, variation points and variants in the variability model of the EPB system are bound to the functional artifacts modeled in SysML through artifact dependency. The functional artifacts provide knowledge of dependencies between system elements.

After importing the variability model and the functional artifacts to the algorithms, the following results are generated: (i) the "Brake Behavior Thresholds" (BBT) variation point is merged into the "Braking Strategy" (BS) variation point; and (ii) the "Force Monitor on Engine Stop" (FMES) and the "Electrical Action Component" (EAC) variation points are merged into the "Architecture Design Alternatives" (ADA).

The above results show that the PLSE process achieves a reduction from 21 variation points down to 18 for the EPB system variability model. In this case study, the performance of the process is limited due to inadequate functional artifacts provided in [28]. It is anticipated that further reduction can be achieved with additional functional artifacts availability. For example, the "Brake Lock" function is likely related to the "Brake Release" function. To avoid conflict decisions made between the brake system and the driver, designs of the "Brake Lock" and the "Brake Release" may require a consistent choice between the "On System Decision" and the "On User Command". As such, according to the process, the "Brake Lock" variation point could then be merged with the "Brake Release" variation point.

The third case study extends the illustrative example used in the previous sections. As shown in Figure 5, an initial variability with hierarchy is produced based on available functional artifacts (Stage 2). Then, by identifying interactions between variants and merging variation points that have a unique interaction, an optimized variability model is derived and depicted in Figure 6. In this case study, the initial 10 variation points as observed in Figure 5 are reduced to 6 variation points as observed in Figure 6.

The last case study is based on another application to an Engine Control System with support from Rolls-Royce. As the engine design is confidential, the original system models and variability models will not be provided. In this case study, following the PLSE process, several SysML functional artifacts are fed into Algorithm 1 as inputs. An initial variability model is generated with 37 variation points. The dependencies between variants, variation points, and functional artifacts as well as the hierarchical structure of the variability model are derived based on available dependencies within the input functional artifacts. Then, Algorithm 2 is further applied to the initial variability model to reduce variation points where possible. The result is that the 37 variation points are reduced to 24 variation points.

The results of the four case studies are summarized in Table I, with a rounded figure of percentage reduction. The first two cast studies demonstrate that our method is generally applicable to various domains in addition to aerospace domain. The other two case studies suggest that the process is scalable for applications to complex systems with large models.

Based on the quantitative evaluations, we have identified two main limitations for our method. Firstly, the performance of the algorithms relies on the availability of functional artifacts. Although Model-based Systems Engineering methodologies have been adopted in the practice of many engineering fields, functional artifacts such as SysML Activity models may not be fully available in the first instance. As such, this can limit the potential usage of the proposed PLSE process. Secondly, the algorithms are much more useful when applied to large scale systems where a considerable number of initial variation points may present. When applied to small system, like the one in Case Study 1, the reduction, though being large in terms of a percentage, only provides marginal return to the user.

## V. CONCLUSION

This paper has presented an end-to-end PLSE process that formalizes product line models and utilizes ROSE concepts to achieve hierarchical variability modeling and variability optimization through variation points reduction. The process features four key stages, two formal definitions and two algorithms.

By using legacy system models, particularly, functional artifacts that are presented in SysML Activity Diagrams, the proposed process generates a three-layer functional artifact using a method that is developed and adopted by Rolls-Royce. Then, based on the definitions introduced, the three-layer hierarchy is preserved through relational transformation to generate a variability model with also hierarchy. Relations between activities in the functional artifact is further utilized to identify dependencies between variants and between variation points. With the hierarchical structure constructed and the dependencies identified, the first three stages are completed. The above steps are semi-automated by the first algorithm.

To optimize the variability model generated, dependency analysis is performed to identify unique interactions. Variation points are then merged when they are dependent via a unique interaction. This reduction mechanism is captured in Stage 4 of the process and is automated by the second algorithm offered in this work.

To assess the applicability and effectiveness of the process, four case studies have been conducted. These case studies cover different engineering domains including Logistics, Automotive, and Aerospace Engineering; and are at different levels of complexity with initial variation points ranging from 5 to 37. The first two case studies (Logics and Automotive) have demonstrated that the process is not restricted to aerospace engineering; and that the reduction mechanism is applicable to situations where hierarchical information and functional artifacts may not necessarily be available. The other two case studies (Aerospace), derived and generalized from Rolls-Royce engine control systems, demonstrated a significant reduction in the number of variation points, with 40% and 35%. The significant reductions are primarily attributed to the availability of functional artifacts. To this end, we conclude that the main limitation of the process is the availability of system artifacts, e.g. functional artifacts presented in SysML Activity Diagrams. However, with the growing popularity of applying Model-based Systems Engineering approaches in traditional engineering, it is anticipated that artifact availability will eventually become irrelevant. The process is also seen to provide marginal returns in the applications to small systems.

Future research will evolve the current process to address different types of variability dependencies that can emerge in a variability model. For example, the «excludes» relation that states two variants cannot both be chosen in a product configuration; and hardware-software interactions that can be dynamic. Finally, as the current work only considers functional artifacts, future work will extend the scope to include other system artifacts such as physical artifacts.

## ACKNOWLEDGMENT

The author would like to thank the Rolls-Royce Controls and Data Services team, who has supported this research. Special thanks also go to PTC for providing a license for the modeling tool Integrity Modeler.

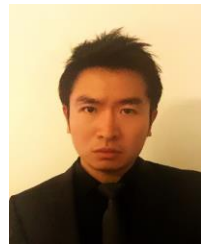

**Mole Li** received his bachelor degree in Computer Science from the University of Hull in 2012 and master degree from Loughborough University in 2013. Currently, he is pursuing his Ph.D. degree at Loughborough University. His current research is in combining Software Product Line Engineering with Model-based Systems Engineering supported by Rolls-Royce Controls group. Mr Li is a member of the OMG and INCOSE.

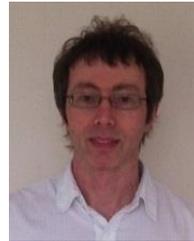

**Alan Grigg** received an honors degree in Mathematics at Thames Polytechnic (now University of Greenwich) in 1985. He subsequently joined British Aerospace (now BAE Systems) and worked for many years on the specification and development of a standardized software architecture for military avionic systems. He worked in the Dependable Computing Systems Centre at University of York where he completed his PhD entitled 'Reservation-based Timing Analysis - A Partitioned Timing Analysis Model for Distributed Real-Time Systems' in 2002. He returned to BAE Systems to work on a mission computer software architecture for the Hawk Advanced Jet Trainer aircraft, then later joined the Systems Engineering Innovation Centre at Loughborough University. In 2012, he joined Rolls-Royce Aero-Engine Controls in Derby to work on Model-Based Systems Engineering and Product Line process improvement programs for Engine Control and Monitoring Systems.

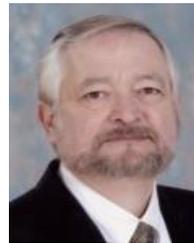

**Charles Dickerson** (M'06–SM'17) received the Ph.D. degree from Purdue University, West Lafayette, IN, USA, in 1980.

He is Chair of Systems Engineering at Loughborough University and a Principal Investigator of complex vehicle system analysis in the Programme for Simulation Innovation. His research and design experience includes MIT Lincoln Laboratory, the Lockheed Skunkworks and Northrop Advanced Systems.

Prof. Dickerson is Chair of the Mathematical Formalisms Group at the OMG and the Assistant Director for Analytic Enablers at INCOSE.

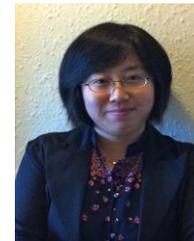

**Lin Guan** is a Senior Lecturer in the Department of Computer Science at Loughborough University. She received her PhD from the University of Bradford, UK, after which she was also appointed as a Research Associate. She then held a project manager/software engineer position in Simulation Systems Ltd shortly, prior to moving to Loughborough.

Her research interests focus on performance modeling/evaluation of heterogeneous computer networks, QoS analysis and enhancements, cloud computing & security, mobile computing, wireless and wireless sensor networks and MBSE (Model-based Systems Engineering). She has published over 80 journal and conference papers and she has been serving as guest co-editor for several international journals, such as those published by Elsevier and Springer. During her PhD, she was awarded the British Federation of Women Graduates Foundation Main Grant in 2004. She then holds two CASE awards, one EngD on EPSRC funded projects and two industrial subcontracts on feasibility study and consultancy. She receives a prestigious award as Royal Society Industry Fellow and EPSRC/LU KTA project.

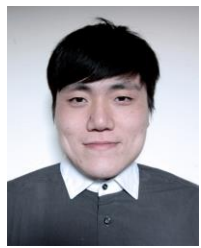

**Siyuan Ji** (M'16) received the Ph.D. degree in physics from University of Nottingham, Nottingham, United Kingdom, in 2015.

He is currently a Research Associate at Wolfson School of Mechanical, Electrical, and Manufacturing Engineering of Loughborough University, United Kingdom. His research focuses on mathematical-based systems engineering methodologies and design algorithms for complex systems and system of systems. Dr. Ji is a member of the IET and a member of the OMG